\documentclass[a4paper,amsmath,amssymb,floatfix,prl,footinbib,twocolumn,superscriptaddress,showpacs,preprintnumbers]{revtex4}

\usepackage{graphicx}
\usepackage{dcolumn}
\usepackage{color}

\newcommand{\comment}[1]{}

\begin{document}

\title{Electron scattering in intra-nanotube quantum dots}

\author{G. Buchs}
\affiliation{EMPA Swiss Federal Laboratories for Materials Testing and Research, nanotech@surfaces, Feuerwerkerstr. 39, CH-3602 Thun, Switzerland}
\affiliation{Kavli Institute of Nanoscience, TU-Delft, P.O. Box 5046, 2600 GA Delft, The Netherlands}
\author{D. Bercioux}
\affiliation{Physikalisches Institut and  Freiburg Institute for Advanced Studies, Albert-Ludwigs-Universit\"at, D-79104 Freiburg, Germany}
\author{P. Ruffieux}
\author{P. Gr\"oning}
\affiliation{EMPA Swiss Federal Laboratories for Materials Testing and Research, nanotech@surfaces, Feuerwerkerstr. 39, CH-3602 Thun, Switzerland}
\author{H. Grabert}
\affiliation{Physikalisches Institut and Freiburg Institute for Advanced Studies, Albert-Ludwigs-Universit\"at, D-79104 Freiburg, Germany}
\author{O. Gr\"oning}
\affiliation{EMPA Swiss Federal Laboratories for Materials Testing and Research, nanotech@surfaces, Feuerwerkerstr. 39, CH-3602 Thun, Switzerland}

\begin{abstract}
Intratube quantum dots showing particle-in-a-box-like states with level spacings up to 200~meV are realized in metallic single-walled carbon nanotubes by means of low dose medium energy Ar$^+$ irradiation. Fourier transform scanning tunneling spectroscopy compared to results of a Fabry-P\'erot electron resonator model yields clear signatures for inter- and intra-valley scattering of electrons confined between consecutive irradiation-induced defects (inter-defects distance $\le 10$~nm). Effects arising from lifting the degeneracy of the Dirac cones within the first Brillouin zone are also observed.
\end{abstract}

\date{\today}



\maketitle

The experimental realization of quantum dots (QDs)~\cite{QD_Leo},
sometimes called ``artificial atoms'', has led to a variety of new concepts in nanotechnology underlying advanced QD-based devices for applications in promising fields like nanoelectronics, nanophotonics and quantum information/computation~\cite{Review_Hanson,Review_Kern,Chuang_book}.
Frequently, for these applications a QD needs to be contacted by source, drain, and gate electrodes. 
In the field of semiconductor heterostructures the excitation energies of contacted QDs are usually so small that the devices can only be operated at cryogenic temperatures. 
A promising candidate for room temperature active dots are intra-nanotube QDs formed within a single-walled carbon nanotube (SWNT) by means of two local defects~\cite{grifoni:2001}. For defect separations of order 10~nm the dot excitation energies are well above 100~meV and thus large compared with $k_\text{B}T$ at room temperature. 
Furthermore, the remaining sections of the SWNT to either side of the confining defects provide natural source and drain electrodes. 
So far, SWNT-based QD prototypes have been realized by tunneling barriers at metal-nanotube interfaces and/or by gate electrodes ~\cite{Review_Christian}.
Several authors have analyzed defect-induced standing waves by means of scanning tunneling microscopy (STM) ~\cite{Lemay_nat01, lieber_PRL, lee:2004}. However, a detailed description of the scattering dynamics of electrons in and out of the QD is absent. Elaborate studies have only been reported for epitaxial graphene with defects, where an analysis of standing waves in Fourier space has permitted to distinguish between contributions to the wave modulation due to inter- and intra-valley scattering~\cite{rutter:2007}.
\begin{figure}[!t]
\centering
\includegraphics[width=0.75\columnwidth]{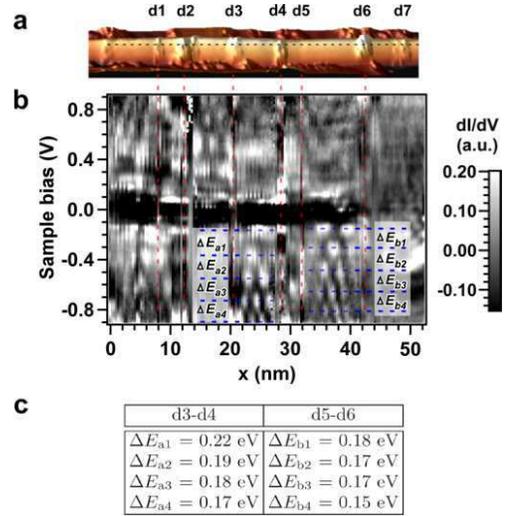}
\caption{(a) 3D topography image \cite{WSXM} of a $\sim$ 50 nm
long portion of an armchair SWNT exposed to 200 eV Ar$^{+}$ ions,
recorded in the constant current mode with a sample-tip bias voltage
($V_\text{s}$) of 1 V (sample grounded) and a tunneling current ($I_\text{s}$) of
0.1 nA, $T = 5.3$ K.  (b)
Corresponding $dI/dV$-scan with background subtraction, recorded
along the horizontal dashed line in (a), $\Delta x =$  0.34 nm.
The $dI/dV$ spectra are recorded through a lock-in detection of a
12 mV rms ($\sim$ 600 Hz) a.c. tunneling current signal added to
the d.c. sample bias under open-loop conditions ($V_\text{s}=$ 0.9
V, $I_\text{s}=$ 0.3 nA). (c) Energy spacings between discrete 
states visible in (b) in the negative bias range
between defect sites d3-d4 ($\sim$ 7.9 nm) and d5-d6 ($\sim$ 9.9
nm).} \label{QBS_1}
\end{figure}
In this Letter we investigate electron standing waves in intra-tube QDs created in SWNTs irradiated with medium energy Ar$^+$ ions. This promising alternative to build intra-tube QDs has been suggested by observations of electronic confinement in metallic SWNTs due to intrinsic defects~\cite{Bockrath01}. We first show that by virtue of this technique it is indeed possible to realize QDs with a level spacing considerably larger than the thermal broadening at room temperature. Then, by means of Fourier-transform scanning tunneling spectroscopy
(FTSTS) combined with a Fabry-P\'erot electron resonator model we are able
to describe the dominant scattering mechanisms and to
identify contributions from inter- and intra-valley scattering.

   Our measurements were performed in a commercial (Omicron), ultrahigh vacuum LT-STM setup at $\sim$ 5 K.
   Extremely pure HipCo SWNTs~\cite{Smalley01} with an intrinsic defect density $<$ 0.005 nm$^{-1}$
   were deposited onto Au(111) surfaces from a 1,2-dichloroethane suspension~\cite{Buchs_NJP_07}. \emph{In situ}
   irradiation with medium energy Ar$^{+}$ ions was performed in a way to achieve a defect density of about
   0.1 nm$^{-1}$~\cite{Buchs_Ar}. Figure~\ref{QBS_1}a shows a 3D STM image of a $\sim50$~nm long portion of
   an armchair SWNT irradiated with $\sim200$~eV ions. Defects induced by medium energy Ar$^{+}$ ions appear
   typically as hillocks with an apparent height ranging from 0.5~{\AA} to 4~{\AA} and a lateral extension
   between 5~{\AA} and 30~{\AA}. 
   We recorded consecutive and equidistant $dI/dV$ spectra
   (proportional to the local density of states (LDOS)~\cite{Tersoff85}) along the tube axis. 
   Typical $dI/dV(x,V)$ data sets, called
   $dI/dV$-scans in the following, consist of 150 $dI/dV$ spectra recorded on topography line scans of 300~pts.
   Figure~\ref{QBS_1}b shows a $dI/dV$-scan with a spatial resolution $\Delta x =$ 0.34~nm recorded along
   the horizontal dashed line drawn in (a), running over seven defect sites (d1-d7). A third order polynomial
   fit has been subtracted from each $dI/dV$ spectrum to get a better contrast. Defect-induced
   modifications in the LDOS are revealed as one or more new electronic states at different energy values,
   spatially localized on the defect sites. First-principle calculations show that medium energy Ar$^{+}$
   ions essentially give rise to single vacancies (SV), double vacancies (DV) and also C adatoms (CAd) on
   SWNTs~\cite{Antti_07}. 
   Based on these results we can confidently assume that the
   created defects in the present work are mainly of vacancy-type.

\begin{figure}[t]
\centering
\includegraphics[width=0.85\columnwidth]{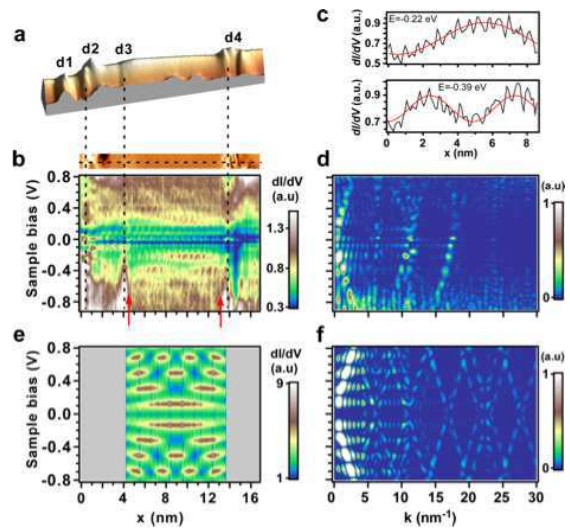}
\caption{(a) 3D STM topography image \cite{WSXM} of a metallic SWNT treated with 200 eV Ar$^{+}$ ions, showing four defects d1-d4. (b) Line-by-line flattened topography image of the tube in (a) including defects d2-d4, with the corresponding $dI/dV$-scan recorded along the horizontal dashed line. $V_\text{s} = 0.8$ V, $I_\text{s} = 0.32$ nA,  $T = 5.21$ K, $\Delta x =$ 0.1 nm. (c) $dI/dV$ line profiles of the first two modes in the negative bias range, recorded between  the red arrows drawn in (b). (d) $\left|dI/dV(k,V)\right|^{2}$ map calculated from the $dI/dV$-scan in (b) between  the red arrows. (e) Differential conductance calculated within the Fabry-P\'erot electron resonator model for a $(7,4)$ SWNT with a defect distance of 9.5 nm, intra- and inter-valley scattering parameters equal to $0.35$ for both impurities, with the corresponding $\left| dI/dV (k,V) \right|^{2}$ map in (f).}
\label{QBS_Fourier}
\end{figure}

   Several broad discrete states characterized by a modulation of the $dI/dV$ signal in the spatial direction
   are observed in the negative bias range between d3-d4 and d5-d6, and in the positive bias range between
   d2-d3. These states show a discrete number of equidistant maxima following a regular sequence $i,\, i+1,\, i+2...$
   for increasing $\left|V_\text{bias}\right|$, similar to the textbook 1D particle-in-a-box model.
   Within this model it is possible to estimate the level spacing around the charge neutrality point
   (CNP) for discrete states observed for example in short SWNTs~\cite{Lemay_nat01,Rubio_prl99}. Assuming a
   linear dispersion $E=\hbar v_\text{F} k$ around the two inequivalent Fermi points $\mathbf{K}$ and
   $\mathbf{K'}$ for a SWNT with finite length $L$, the energy spacing is then given by:
   \begin{equation}
       \Delta E = \hbar v_\text{F} \frac{\pi}{L} = \frac{h v_\text{F}}{2L} \simeq \frac{1.76}{L}\, \text{eV} \cdot \text{nm}
       \label{deltaE}
   \end{equation}
   with $L$ in nm and the Fermi velocity $v_\text{F}= 8.5 \cdot 10^{5}$~m$\cdot$s$^{-1}$~\cite{Lemay_nat01}.
   The energy spacings $\Delta E_\text{a1}$-$\Delta E_\text{a4}$ in the negative bias range between d3-d4 and
   $\Delta E_\text{b1}$-$\Delta E_\text{b4}$ between d5-d6 are reported in Fig.~\ref{QBS_1}c. Using the sequence
   of maxima we can determine the level spacing closest to the CNP: $\Delta E_\text{a1} = 0.22$~eV [$\Delta E_\text{b1} = 0.18$~eV] between d3-d4 [d5-d6]. This corresponds to a defect distance of $L = 8$~nm [$L = 9.78$~nm] for the defect
   separation d3-d4 [d5-d6], in good agreement with the measured value at the center of the defect sites
   $L \simeq 7.9$~nm [$L \simeq 9.9$~nm]~\cite{note:one}.
   These results show artificial defect-induced electron confinement regions in
   metallic SWNTs, \emph{i.e.} intratube QDs. Importantly, spatially close defects can be generated with our
   method, allowing level spacings which are much larger than the thermal broadening at room temperature of
   $k_\text{B} T \simeq 25$~meV.

   Figure~\ref{QBS_Fourier}a shows a $\sim$16~nm long section of a metallic SWNT exposed to 200~eV Ar$^{+}$
   ions with four defect sites (d1-d4). A line-by-line flattened topography image of the same tube
   between defects d2-d4 is displayed in panel (b) with the corresponding $dI/dV$-scan recorded
   along the horizontal dashed line. Two discrete states are clearly visible in the negative bias range between
   d3 and d4, at energies $E = -0.22$ eV and $E = -0.39$ eV. The measured energy spacing of about 170 meV fits
   well with the value of 177 meV obtained from Eq.~(\ref{deltaE}) for a defect separation of about 9.5~nm.
   Line profiles of $dI/dV$ signals recorded between the drawn red arrows in (b) and displayed in (c) show a 
   clear oscillatory behavior characterized by a rapid oscillation with an average
   wavelength of 0.7~nm modulated by a slower variation of the amplitude. This slow modulation, which shows a
   decreasing wavelength for increasing $\left| V_\text{bias} \right|$, has been fitted with the function
   $\left| \psi \left( x \right) \right|^{2}=A+B \sin \left( 2 k x + \phi \right)$, where $\phi$ is an arbitrary
   phase and the factor 2 originates from the fact that $\left| \psi \left( x \right) \right|^{2}$ is probed.

   More details on the observed oscillatory behavior are obtained by means of FTSTS, where line-by-line 
   Fourier transforms are performed on the $dI/dV$-scan in Fig.~\ref{QBS_Fourier}b,
   between the positions indicated by the red arrows. From the resulting $\left| dI/dV(k,V) \right|^{2}$ map in (d),
   we observe that the Fourier spectrum of each discrete state is composed of several components~\cite{note:two}. 
   Whereas the individual
   low frequency peaks between $k=0$ and $k=4$ nm$^{-1}$ with a high intensity for each discrete state correspond to
   the slow modulation discussed above, the rapid oscillation in $dI/dV$ is produced by several components around
   $k=11$~nm$^{-1}$ and $k=17$~nm$^{-1}$. These Fourier components are aligned along sloped lines, indicating the
   energy dispersive nature of these features. Around $k=17$~nm$^{-1}$, a unique positively sloped line is clearly
   visible, with $dE/dk \simeq$ 0.32~eV\,nm, whereas two lines with positive and negative slopes can be distinguished
   around $k=11$~nm$^{-1}$, with a measured slope of about 0.3~eV\,nm.

\begin{figure}[t]
\centering
\includegraphics[width=0.85\columnwidth]{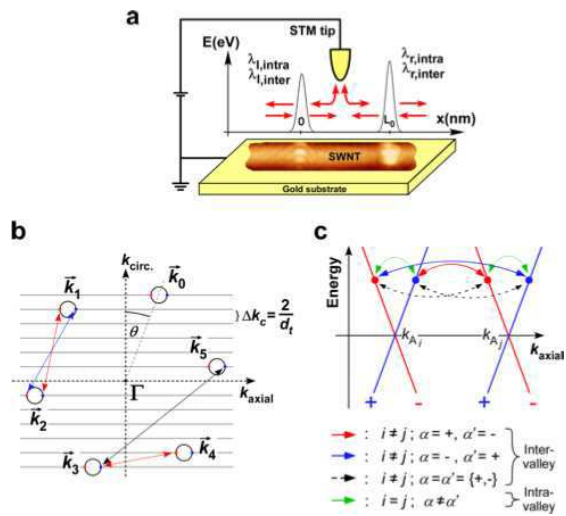}
\caption{(a) Outline of the experimental set-up and of the
Fabry-P\'erot resonator model. (b) Sketch  of the graphene valleys
at a fixed energy for a $(7,4)$ metallic SWNT. The parallel
horizontal $\textbf{k}$-lines are reminiscent from the
quantization in the circumferential direction. (c) Sketch of the
1D-scattering processes between two nonequivalent valleys:
intra-valley (green arrows) and inter-valley scattering
(red, blue and black arrows). The processes indicated by
black arrows are only relevant if a commensurability condition is
fulfilled. $k_{\text{A}_{i}}$ and $k_{\text{A}_{j}}$ are the CNP
axial momenta components of valleys $i$ and $j$, respectively.
\label{fig3}}
\end{figure}

\begin{figure}[t]
\centering
\includegraphics[width=0.65\columnwidth]{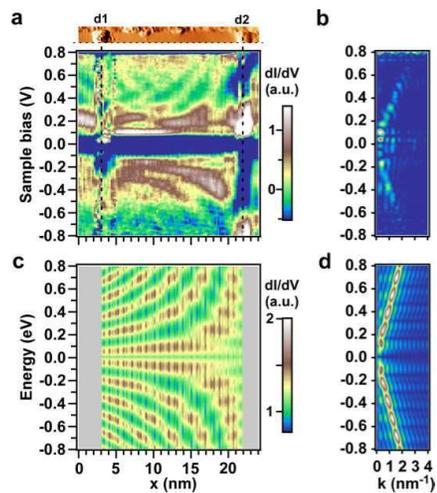}
\caption{(a) STM current error image \cite{WSXM} of a metallic SWNT with two defect sites d1 and d2 produced by an exposition to 1.5 keV Ar$^+$ ions, with the corresponding $dI/dV$-scan recorded along the horizontal dashed line. 
Interference pattern visibility is improved via a background subtraction.
 (b) Corresponding $\left| dI /dV (k, V ) \right|^{2}$ map limited to low frequencies contributions. $V_\text{s} =$ 0.8 V, $I_\text{s} =$ 0.3 nA, $T = 5.3$ K, $\Delta x =$ 0.22 nm. (c) Calculated $dI/dV$-scan for a $(10,7)$ SWNT with a length of $18.5$ nm corresponding to the average distance between the defects d2 and d3. The intra- and inter-valley impurity strengths are equal to $0.025$ and $0.05$ for the left and the right impurity, respectively. (d) Corresponding $\left| dI /dV (k, V ) \right|^{2}$ map limited to low frequencies contributions.
\label{fig4}}
\end{figure}

   In order to fully explain the experimental features, we use a Fabry-P\'erot electron resonator model (see
   Fig.~\ref{fig3}a) considering interference at fixed energy of electron states scattered by impurities.
   These states can be easily identified considering an unrolled SWNT, \emph{i.e.} a graphene sheet showing
   periodic strings of defects along the circumferential direction. The impurities break the translational
   invariance along the SWNT axis, allowing low energy electron scattering among the six valleys or
   \emph{Dirac cones} of the first Brillouin zone. The momenta exchanged in these processes can be decomposed
   in axial $k_\text{A}$ and circumferential $k_\text{C}$ components with respect to the tube axis. The former
   give rise to the interference pattern resulting in the standing waves, whereas the latter modulate the
   intensity of the standing waves in a non-linear way, \emph{i.e.} a larger $k_\text{C}$ component leads to
   a lower intensity. In the calculated $\left| dI/dV(k,V) \right|^{2}$ maps, these standing waves give rise to
   intensities at $k$-values corresponding to the axial component $k_\text{A}$. Therefore, the
   $\left| dI/dV(k,V) \right|^{2}$ maps show a weighted projection of the 2D space of possible scattering
   vectors along the axial direction. The situation is depicted in Fig.~\ref{fig3}b. Two distinct scattering
   mechanisms take place: \emph{intra}- and \emph{inter}-valley scattering. For the first process within the
   same valley the momentum exchange is small, even zero at the CNP (green process in (c)); the second
   process connects different valleys (blue, red, and black processes in (b) and (c)). Both scattering
   mechanisms are related to the presence of SVs and DVs \cite{ando05}.

   This analysis reduces the scattering processes to a series of weighted 1D scattering events among electrons
   with a linear energy dispersion and axial momenta $k_{\text{A}_i},k_{\text{A}_j}$ (see Fig.~\ref{fig3}c).
   We model the impurities  as  delta-like potentials placed at a distance $L$, and the STM tip is included
   by allowing electron tunneling to an external electrode~\cite{bercioux:2009,note:coulomb} (see Fig.~\ref{fig3}a).
   Figure~\ref{QBS_Fourier} shows a comparison between the measured (b) and the calculated
   (e) LDOS for the case of a SWNT with two identical impurities. The measured SWNT has a chiral angle
   $\theta \approx 21^\circ$ and shows three dispersion lines at $k=6.1,\, 10.7$ and $16.8$~nm$^{-1}$,
   compatible with a $(7,4)$ metallic SWNT. The numerically evaluated $\left| dI/dV(k,V) \right|^{2}$ map
   shown in (f) unveils richer structure than the experimental one. These differences can be attributed
   to the finite resolution of the tip.
   The components centered around $k=0$~nm$^{-1}$ are more intense than the others because they are associated
   with intra-valley scattering occurring at all six valleys. In the measured $\left| dI/dV(k,V) \right|^{2}$
   the dispersion lines around $k=10.7$ and $16.8$~nm$^{-1}$ show a more intense signal for the positive slope
   branch than for the negative one. For the component centered at 16.8 nm$^{-1}$, the negative slope branch
   is almost missing. Similar behavior has been observed in all samples investigated.
   For armchair SWNTs, this effect has been related to an interplay between symmetry properties of defects and electronic bands resulting in a suppression of $\pi\to\pi$ scattering~\cite{lieber_PRL}.
However, in our case of chiral SWNTs the relation between the parity of the $\pi$- and $\pi^*$-band is more complex and there is no obvious explanation of the observed branch asymmetry. Since our observations are of  pivotal importance to the electric transport properties of real SWNT devices, this issue certainly deserve further in depth experimental characterization and theoretical explanation.

   If the commensurability condition $L= m \pi/\Delta k_\text{A}$
   with $m$ integer and $L$ the defect separation is not fulfilled, inter-valley scattering takes place only
   between electrons with opposite direction of motion and implies an asymmetry of the spots in the $\left| dI/dV(k,V) \right|^{2}$ along the positive and negative slop branches. Contrarily,
   intra-valley scattering always fulfils this condition with $m=0$, therefore showing symmetric spots around $\Delta k_\text{A}=0$ (see Fig.~\ref{fig3}(c)).

   Figure~\ref{fig4}a shows a current error image of a metallic SWNT which has been exposed to 1.5 keV
   Ar$^{+}$ ions, and the corresponding $dI/dV(x,V)$-scan recorded along the tube axis through two defect sites
   labeled d1-d2. Here, instead of clear ``textbook-like'' modes as shown in Fig.~\ref{QBS_Fourier}, we observe
   curved stripes in the interference pattern between defects d1-d2. However, the corresponding $\left| dI/dV(k,V) \right|^{2}$
   on the right hand side clearly shows well-defined small momentum spots with an average energy separation of about
   90 meV in good agreement with Eq.~(\ref{deltaE}) giving $\Delta E \simeq 95$ meV for a measured defect
   separation $L$ of about 18.5 nm. These features are also captured by the Fabry-P\'erot electron resonator model
   if the two impurities have different scattering strengths $\lambda$, as shown for the simulated $dI/dV$-scan with $\lambda_\text{L}=0.05$ and $\lambda_\text{R}=0.15$ in Fig.~\ref{fig4}b. There is now a clear asymmetry characterized
   by stripes showing an increasing curvature when moving from the weaker left to the stronger right impurity.
	There can also be energy dependent differences in the scattering strengths of the defects leading to energy dependent asymmetries in the standing wave pattern as seen in Fig.~\ref{QBS_1}. 
The choice of ion energy can significantly change the type of defects produced and could therefore potentially be used as a parameter to control to some extend the scattering configuration~\cite{Buchs_Ar}.
	
In summary, studying intratube QDs in SWNT by FTSTS in combination with simulation, we provided an analysis of the dominant electron scattering processes. Clear signatures for inter- and inta-valley scattering were observed, and scattering effects arising from lifting the degeneracy of the Dirac cones were identified. The here applied strategy will be useful to investigate the scattering properties of other local modifications of SWNTs like e.g. chemical functionalities.

We thank M.~Grifoni, O.~Johnsen, S.~G.~Lemay, T.~Nakanishi, Y.~Nazarov, D.~Passerone, C.~Pignedoli, and in particular Ch.~Sch\"onenberger  for fruitful discussions. This work was supported by the Swiss National Center of Competence in Research MANEP,  and the Deutsche Forschungsgemeinschaft (DFG).

\vfill
\break


\begin{thebibliography}{99}
\bibitem{note:one} The slight decrease of $\Delta E$ for increasing $\left| V_\text{bias} \right|$ can be attributed to the
divergence from linearity of the real dispersion relation for energies far from the CNP~\cite{ando05}.
\bibitem{note:two} To avoid aliasing effects, we ensure that the highest spectral component fulfills the Nyquist sampling theorem.
\bibitem{note:coulomb} Coulomb interaction in the SWNT is partially screened by the gold substrate and is not taken into account explicitly.
\end{thebibliography}
\end{document}